\newif\ifAMStwofonts
\AMStwofontstrue 
\def\vgrad{{\bf \nabla}}

\def\rmin{{r_{_{\rm min}}}} 
\def\rmax{{r_{_{\rm max}}}} 
\def\rz{{r}} 
\def\vh{{v_{_{\rm h}}}} 

\def\Pm{{P_{_{\rm m}}}} 
\def\ti{{t_{_{0}}}} 
 
\def\Qks {{Q_{_{\rm KS}}}} 
\def\rcu{ {r_{_{\rm cut}} }}
\def\md{ {M_{_{\rm disk}} }}
\def\psid{ {\Psi_{_{\rm disk}} }}

\def\kpc{ {\; {\rm kpc}}}
\def\Ns{N_{\rm s}}

\def\vv{\bf v}

\def\gsim{~\rlap{$>$}{\lower 1.0ex\hbox{$\sim$}}}

\def\td{t_{_{\rm d}}}
\def\tobs{t_{_{\rm obs}}}

\def\ltsim{\lower.5ex\hbox{$\; \buildrel < \over \sim \;$}}
\def\gtsim{\lower.5ex\hbox{$\; \buildrel > \over \sim \;$}}
\def\ltsim{\lower.5ex\hbox{$\; \buildrel < \over \sim \;$}}
\def\gtsim{\lower.5ex\hbox{$\; \buildrel > \over \sim \;$}}

\def\kms{\mbox{km\,s$^{-1}$}}

\def\dd{\,{\rm d}}

\newcommand{\beq}{\begin{equation}}
\newcommand{\eeq}{\end{equation}}
\def\beqa{\begin{eqnarray}}
\def\eeqa{\end{eqnarray}}
\def\la{\lower.5ex\hbox{$\; \buildrel < \over \sim \;$}}
\def\ga{\lower.5ex\hbox{$\; \buildrel > \over \sim \;$}}
\def\fixit#1{}

\def\dd{{\rm d}}
\documentclass[preprint2]{aastex}
\usepackage{natbib}
\begin{document}

\title{ Ergodic Considerations in  the  gravitational potential of the Milky Way }

\author{Adi Nusser}
\affil{Physics Department and the Asher Space Science Institute- 
Technion, Haifa 32000, Israel}

\email{adi@physics.technion.ac.il}

\slugcomment{Accepted to ApJ}

\begin{abstract}
A method is proposed for constraining the Galactic gravitational
 potential from  high precision observations of the phase space coordinates of 
a system of relaxed tracers. The method relies on an ``ergodic'' assumption that 
the observations
are representative of the state of the system at any other time.
The observed coordinates serve as  initial conditions  for  
moving  the tracers forward in time in an assumed model for the gravitational field.  
 The validity of the model is assessed  by the  
    statistical equivalence  between the observations and the 
  distribution of tracers at randomly selected times.
The applicability of this ergodic method is not restricted by any 
assumption on the form or symmetry of the potential.  However, it requires 
high recision observations  
 as those that will
be obtained from missions like SIM and GAIA.

\end{abstract}

\section{Introduction}
Measurements of velocities of cosmological objects are a classic probe of the 
mass distribution 
on all scales.
  It was the motions of individual galaxies in galaxy clusters which
 first showed that luminous mater contributed only a small fraction to the total mass in
 clusters \cite[]{Zwicky37}, implying the existence of dark matter. 
 The combined mass of the Milky Way (or the Galaxy)  and M31 is 
constrained from the observed relative  motion between the two galaxies and the 
requirement that 
the initial distance between their respective centers of mass vanishes near the 
Big Bang \cite[]{KW59}. 
Line-of-sight velocities of other galaxies in the Local Group  of galaxies  
also are used to estimate its mass by the condition of vanishing initial
 distances \cite[]{Peebles89}. 
On scales 10s of Mpcs, peculiar motions (deviations of Hubble flow) of galaxies 
  constrain the global mass 
density in the Universe \cite[e.g.][]{Nusser08}.

Determining the mass distribution in the Milky Way
is particularly important. 
There is ample information on the  baryonic content  of the Galaxy which could 
be modeled in detail
only if the dark matter distribution is known. 
The rotation curve of the Galaxy is limited to distances smaller than 20 kpc and
 does not provide any 
information on deviations from spherical symmetry of the halo.  The
 mass distribution at larger 
distances,  
motions of Galactic satellite galaxies, globular clusters and stars are invoked
\cite[e.g.][]{sakamoto03}.
 Constraints on the Galactic mass from these tracers are 
  derived from  the condition that the  observed speeds of Galactic objects 
   do not exceed the escape velocity. 
  This approach yields 
  only a lower limit and 
 is mostly  sensitive to the  highest velocity objects. The vast majority of the sample
  objects play no role in 
 deriving the mass limit.  Alternatively, one could adopt a Bayesian likelihood 
 formalism in which the 
 phase space distribution function is assumed to follow certain form 
which could be matched with the observations to 
probe the Galactic potential field  \cite[e.g.][]{LT87,Koch96,WE99} .
 
Proper motions, resulting from  velocities perpendicular to the line-of-sight, are currently
 measured only for nearby 
tracers \cite[e.g.][]{sakamoto03}. 
Therefore, most current mass estimates rely on the measured
 line-of-sight motions of 
individual components.  During the next few years, accurate proper motions for 
a large sample
 of Galactic tracers are expected to be measured by 
the space missions {\it Global Astrometry Interferometer for Astrophysics }
 (GAIA) \cite[]{LP96} and 
{\it Space Interferometry Mission }
(SIM) \cite[]{rr1}.
Even accurate phase space information require additional assumptions in order to 
constrain the Gravitational potential \footnote{The gravitational force field is 
equal to the acceleration rather than velocities. 
 Measuring the acceleration of a tracer with orbital period $\td$ over an observing 
 time $\tobs$ requires astrometry 
 with  angular resolution a factor $\tobs/\td$ higher than the precision needed for 
  velocities. Since
 $\tobs\sim$ a few years  while $\td\sim$  a few Gyrs, the task is out of reach in the
  near future. }. 
We present here a general method which 
relies 
on high precision measurements of positions and velocities  of  tracers.
The method assumes that the tracers 
 have reached  an equilibrium  state in the Galactic gravitational field.

\section{The method}
\label{sec:method}
The expected precision of future data of proper motions, radial velocities and distances 
will allow an accurate determination of the  orbits of tracers   in a given 
 Galactic gravitational field. 
This motivates the ``ergodic" method which constrains the gravitational potential by the hypothesis
 that the state of a dynamically relaxed 
system of tracers at any time 
is statistically  representative of  that system at any other time. 
 The method is described  as follows.
\begin{enumerate}
\item assume a model for the gravitational potential field  
$ \Psi(r,\alpha,\beta\dots)$ 
where  $\alpha$, $\beta $ etc are  free parameters.
\item for a given choice of free parameters, use the observations as initial 
conditions at time $t=\ti$ to advance the 
particles (tracers),  in the gravitational field $\Psi$, for sufficiently long time. 
\item select snapshots of the particle distribution at times $t>\ti$ 
\item compare between statistical measures of the particle distributions in the observed
 data ($t=\ti$) and 
in the snapshots ($t>\ti$). If necessary, 
repeat (ii)-(iv) with a different choice of free parameters until  a reasonable 
agreement is reached.
\end{enumerate}
We work  here with four statistical measures.
The first is the $N$ statistic computed as follows. 
Consider all snapshots (at $t>\ti$) corresponding to an assumed value of $\alpha$. 
For a snapshot $s$, we  compute $\Ns$ defined as the number   of particles with Galactic distances 
that are larger than the respective observed distances.
We define  the statistic $N$ as $N=<\Ns>$,  the average of $\Ns$ over all snapshots.
 We further define  
 $\sigma=<(\Ns-N)^2>^{1/2}$, as the r.m.s. scatter in $\Ns$.   
In the limit of a large number 
of tracers,  $n$,  the limits $\Ns=n/2$ and $N=n/2$ are  approached for the correct model. 
The second is the distribution function of the Galactic distances, $\rz$, of tracers, i.e. 
the density of tracers as a function of 
distance. 
 The third measure is distribution function of 
  \begin{equation}
 \xi=\frac{\rz-\rmin}{\rmax-\rmin}\; ,
    \label{eq:xi}
\end{equation}
computed from the Galactic distances, $\rz$, of particles  in each snapshot. 
 The pericenter, $\rmin$,  and apocenter, $\rmax$, of each particle is computed from the 
 numerically integrated orbits.
The distributions of $r$ and $\xi$  computed from the observations given at 
  $\ti$ should be statistically equivalent to the respective distributions 
in any snapshot at $t>\ti$ if the system is evolved to $t>\ti$ using a gravitational potential which is a
 reasonable approximation to the true potential field.
Significant differences between the initial and later  distributions should appear if 
the   integration at $t>\ti$ is performed  with sufficient deviations from the original
 gravitational potential.  
To see this  consider the simple case of  
  particles executing circular motions in the gravitational field of a 
point mass, $M_0$. Now let us use a snapshot of the motion of this system at time $t=\ti$ as 
initial conditions to move the particles forward in time 
in the field of the point mass  $M\ll M_0$. The pericenters, $\rmin$,  of orbits obtained  with $M$ will 
equal  the  initial distances so that $\xi = 0$ for all particles at time $\ti$.
However, the distances at a randomly selected later time $t\gg \ti$ span 
 the range $\rmin$ to $\rmax$, yielding $\xi$ from 0 to 1 (the $\xi$ distribution
peaks at the end points where the radial velocities are zero). 
Therefore, $M\ll M_0$ could be rejected by the gross mismatch between the
$\xi$ distributions  at the initial time $\ti$ and at $t\gg \ti$. Similar arguments apply
 for the distribution function of 
$r$.

The fourth statistic is based on the virial relation between the total kinetic energy
 (per unit mass)  $T=\sum \dot {\bf r}^2/2$ 
and a term $ W$ related to the gravitational force field. In order to derive the explicit 
relation, we perform a scalar multiplication of the radius vecor, ${\bf r}$  
of each particles with 
the corresponding equation of motion  $ \dd {\vv} /\dd t=-{\mathbf \nabla}\Psi $. 
 Summation over all particles and time averaging gives
\begin{equation}
2\overline T=\overline { W }\; , \quad {\rm where} \quad  W\equiv 
 -\sum_{tracers}{\bf r}\cdot {\bf \nabla}\Psi \; ,
\label{eq:vir}
\end{equation}
and the over-lines denote time averaging. Therefore, as our fourth statistic we consider
the ratio $|W|/2T$ computed directly from the observations for an assumed form of $\Psi$. 
Since the virial relation applies to time average quantity, this statistic, computed at
 a single time, 
is expected to deviate from unity for a finite number of particles. 
Therefore, although the statistic  could be  computed without evolving the particles 
in time, snapshots at  later times will be used to compute the scatter  in the statistic. 
This ratio differs from the other statistics in that it directly involves the
 observed  velocities. 
It is only  sensitive to the force field at the observed positions of the tracers.
i\section{Tests}
Thorough tests of the performance of the method should take into account the 
details of future observations. 
Although this is necessary in order to provide reliable error estimates, 
the task seems 
futile at this early stage. The  preliminary tests presented here aim at  
demonstrating  the ability of the method  to yield significant constraints on 
 a one parameter family 
for the form of $\Psi$ as given below. 
  We test the method using  catalogs of mock data of dynamically relaxed tracers.
The system is assumed to exist in a galaxy  made of a spherical dark matter halo and 
a baryonic  disk, neglecting  the gravitational effects of a bulge.
The gravitational potential of the  halo is taken as \cite[]{sakamoto03,dinescu99}
\begin{equation}
\Psi_{_{\rm halo}}= \left\{
\begin{array}{ll}
\vh^2\ln\left[1+(r/d)^2\right]    -\Psi_0 \; ,& {\rm for} \quad r<\rcu\\
 -2\frac{\vh^2}{r} \frac{\rcu^3}{\rcu^2+d^2}\; , & {\rm otherwise}\; ,
\end{array}\right.
\label{eq:poth}
\end{equation}
where $\Psi_0$ is a constant ensuring  the continuity of the force field per unit mass, $-\vgrad\Psi_{_{\rm halo}}$, at
$r=\rcu$. We take $\vh=128 \kms$, $d=12\kpc$ and 
$\rcu=170\kpc$ as in \cite{sakamoto03}. 
For the potential of the disk we work with the form (Miyamoto \& Nagai 1975),
\begin{equation}
\label{eq:potd}
\psid= \frac{-G\md }  {\sqrt{ R^2 +\left(a+ \sqrt{z^2+b^2}\right)^2  }    }\; ,
\end{equation}
where $d=12 \kpc$, $\md=10^{11} M_\odot$, $a=6.5\kpc$ and $b=0.26\kpc$ 
\cite[]{dinescu99}.
  The mock observations are generated as follows. Particles are placed in  the halo 
 by a random Poisson sampling of an underling number density $n(r)\propto 1/r^2$.
 The  corresponding  velocities
are selected randomly from a gaussian distribution  with zero mean and a 
r.m.s proportional to the local escape speed such that  the 
total kinetic energy is half the   absolute value of the potential energy, as implied by 
the virial theorem. 
The equations of motion of the system are  then solved numerically for  
 $10 \rm \; Gyrs$ to ensure dynamical 
relaxation in the gravitational field given in  Eqs.~\ref{eq:poth} \& \ref{eq:potd}.
Outputs of particle positions and velocities obtained from this procedure 
are then identified as the mock catalog  to be used for testing the method.  
As our model gravitational potential we use the  forms in Eqs.~\ref{eq:poth} \& \ref{eq:potd}
 but with 
 $\md$ and $\vh^2$ scaled by a factor
 $\alpha$ according to $(\md, \vh^2)\rightarrow (\alpha \md,\alpha \vh^2)$. 
 The tests are performed 
for one free parameter, $\alpha$.
Following the scheme above,  the mock catalog provide  the initial conditions at time  
$\ti$ for moving the
particles 
for another $10 \rm \; Gyrs$ for an assumed value of $\alpha$.  
Snapshots of particle positions are then tabulated   at 1000  uniformly distributed time 
steps between $\ti$ and $\ti+10\; \rm Gyrs$.
This is done for 17 values of $\alpha$ spanning the range $0.7-1.5$ linearly. The tests will
 demonstrate that 
the method is able to recover the correct value, $\alpha=1$, within an acceptable uncertainty. 

\subsection{Tests with zero measurement  errors}
\begin{figure*} 
\centering
\includegraphics[  scale=0.4,angle=00]{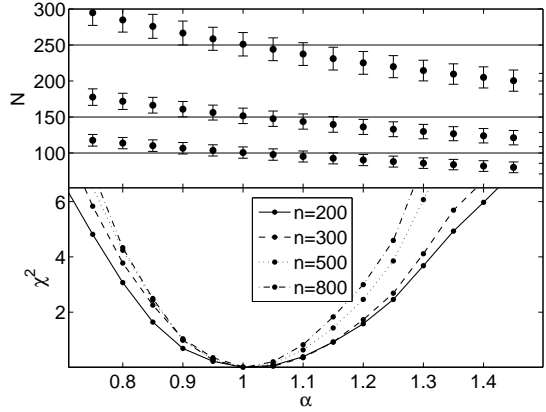}
\vspace{0pt}
\caption{{\it Left:} The statistic $N $ as a function of $\alpha$ shown as the filled circles 
with error-bars representing the r.m.s scatter, $\sigma$ (see text). Three values of the number of tracers are considered, $n=200$,
 300 and 500 and the three horizontal lines indicate corresponding  limiting values $N=n/2$. 
{\it Right:}  Curves of  $\chi^2=(N-n/2)^2/\sigma^2$ as a function of $\alpha$ for several values of $n$, as indicated in the figure.  }
\label{fig:N}
\end{figure*}

We start with the statistic $N$ computed from snapshots corresponding to an assumed $\alpha$.
The filled circles in the left panel of Fig.~\ref{fig:N} show $N$ 
as a function of    $\alpha$, for three values of the number of tracers, $n=200$,  300 and 500. 
The error-bars represent the r.m.s scatter, $\sigma$. 
The horizontal solid lines 
indicate the  limit, $N=n/2$, corresponding to the three values of $n$. For $\alpha$ close 
to  unity,  $N$ is indeed near $n/2$  
even for $n=200$. 
For better quantification of these results we plot in the right panel 
the quantity $\chi^2(\alpha)=(N-n/2)^2/\sigma^2$. Curves of $\chi^2$ 
become narrower as
$n$ is increased, but already with $n=200$ significant constraints on $\alpha$ could 
be derived.  Note that $<(\Ns-n/2)^2>/\sigma^2=1+\chi^2$ so that $\chi^2=3$ correspond to 
 a $2\sigma$ deviation of  $\Ns$. 
 We see that $n=200$ already constrains  $\alpha$ to better than  25\% at the 
 $2\sigma$ level.

\begin{figure*} 
\centering
\includegraphics[scale=0.4 ,angle=00]{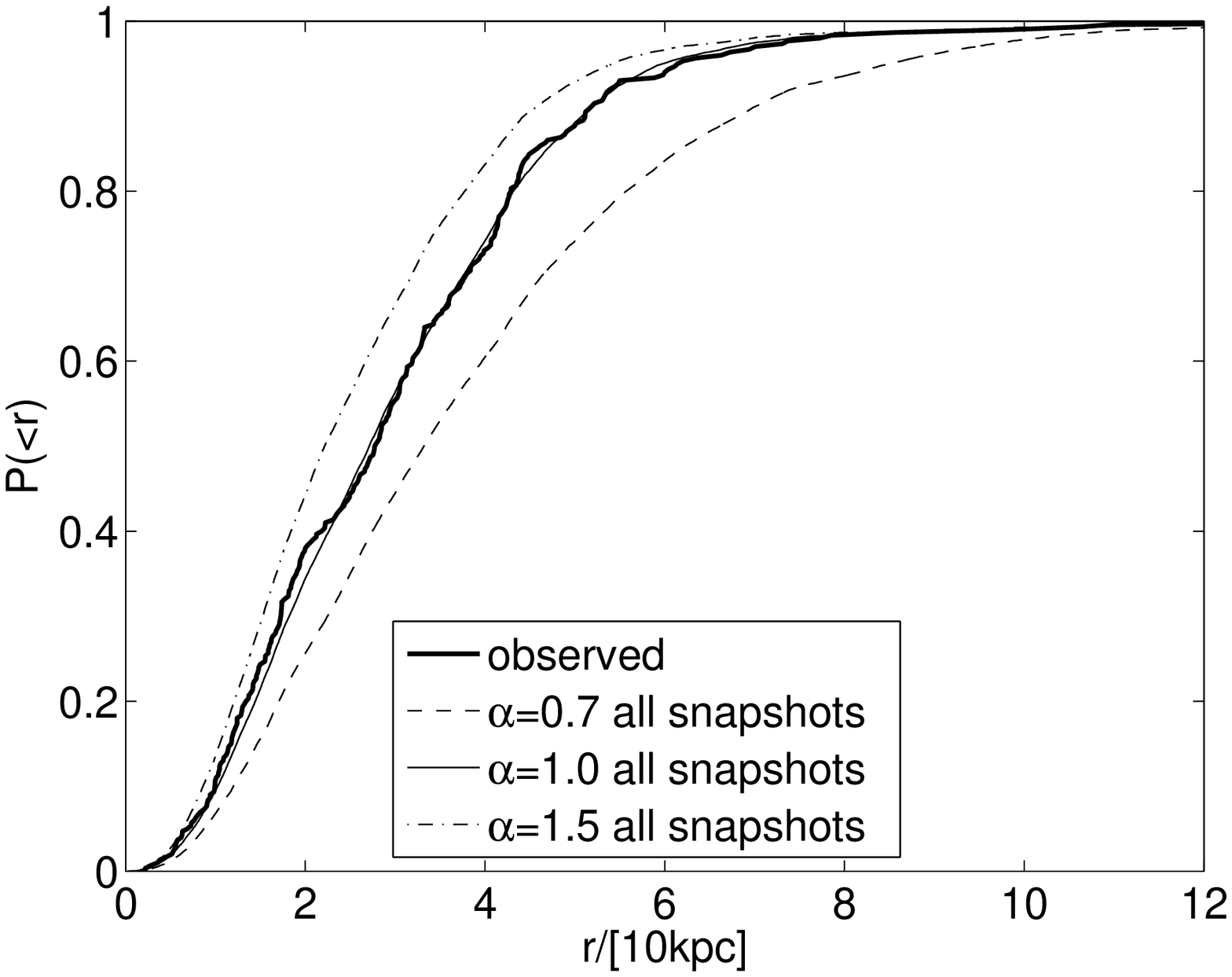}
\includegraphics[scale=0.4 ,angle=00]{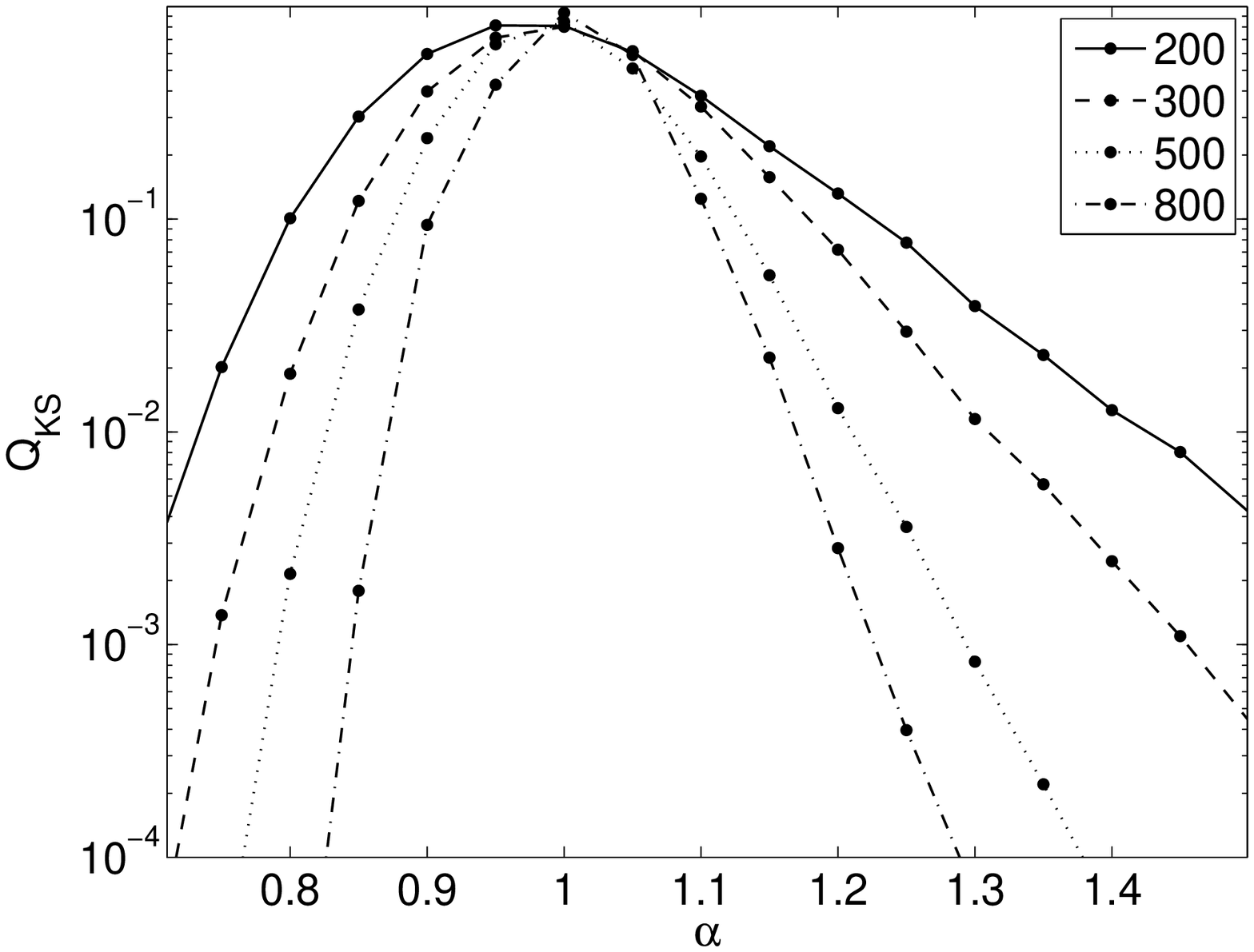}
\vspace{0pt}
\caption{{\it Left:} CPDFs of the distances for $300$ tracers. 
The thick line shows $P(<r)$  
 computed from the observed ``distances:  at $t=\ti$, while  the three  smooth thin lines
 represent the model CPDF $\Pm(<r)$ for three values of $\alpha$, as indicated in the figure.
 {\it Right:} The confidence  level $\Qks$ that $P(<r)$ and $\Pm(<r)$ represent the same 
 distribution, as a function of $\alpha$.}
\label{fig:pr}
\end{figure*}

We now turn to 
 the distribution
function of $r$ and $\xi$ as described in \S\ref{sec:method}.
 We define the cumulative distribution function (hereafter CPDF), $P(<y_1)$,  at any time, 
 as the fraction of particles having $y<y_1$, where $y$ is either $r$ or $\xi$.
 Hereafter, we will treat the CPDF obtained from the ensemble of 1000 snapshots as
  the underlying 
actual model  CPDF and denote it by  $\Pm$. 
In the left panel of Fig.~\ref{fig:pr} we illustrate the differences between the various  CPDFs obtained 
with $300$ tracers.
The thick solid curve is  the CPDF $P(<r)$ obtained from the ``observed" positions in the  
mock catalog, while  the  dashed, thin solid, and dot-dashed lines correspond to  $\Pm(<r)$ for 
$\alpha=0.7$, 1, and 1.5, respectively.  The CPDF $\Pm(<r)$ is sensitive to $\alpha$.  The value  
$\alpha=0.7$ broadens  the distribution of particles towards  larger radii, relative to the ``observed" CPDF, 
while  $\alpha=1.5$ concentrates the particles nearer to the center. 
The CPDF   $\Pm(<r)$ for $\alpha=1$ (smooth thin solid line)  seems consistent with the 
 ``observed" CPDF (thick solid).  
To quantify the  differences between model  and observed ``observed" CPDFs, 
we compute the Kolmogorov-Smirnov (KS)  ``distance" $D_\alpha=\sup_{_r}|\Pm(<r)-P(<r)|$
where the  dependence on $\alpha$ is only through
 $\Pm(<r)$.  Given $D_\alpha$ and the number of tracers, $n$,  we compute the 
significance level, $Q_{_{\rm KS}}$,  of   
the hypothesis that  $P(<r)$ 
and  $\Pm(<r)$ represent the same distribution. 
In the right panel of Fig.~\ref{fig:pr}, we plot  $Q_{_{\rm KS}}$ as a function of $\alpha$ for 
 $n=200$, 300, 500 and 800,  as indicated in the figure.  
 All curves 
peak near the  value, $\alpha=1$. The  only uncertainty in determining $\alpha$ is due
 to the finite number 
of tracers,  hence the  peaks are narrower  for larger $n$. For all values of $n$ considered here, 
the choice $\alpha=1$ 
is never ``rejected" 
at more than the $32\% $ confidence level (i.e. $\Qks<0.32$), equivalent  to a  $1\sigma$ level 
for a normal distribution.  Thus the method produces unbiased estimates of 
 $\alpha$ within the $1\sigma $ errors.  For $n=200$, 
 deviations of $\alpha$ larger  than $\sim 30 \% $ from unity are rejected at
  more than  the  $2\sigma $
 level (i.e. $\Qks<0.045$).
 
\begin{figure*} 
\centering
\includegraphics[   scale=0.4 ,angle=00]{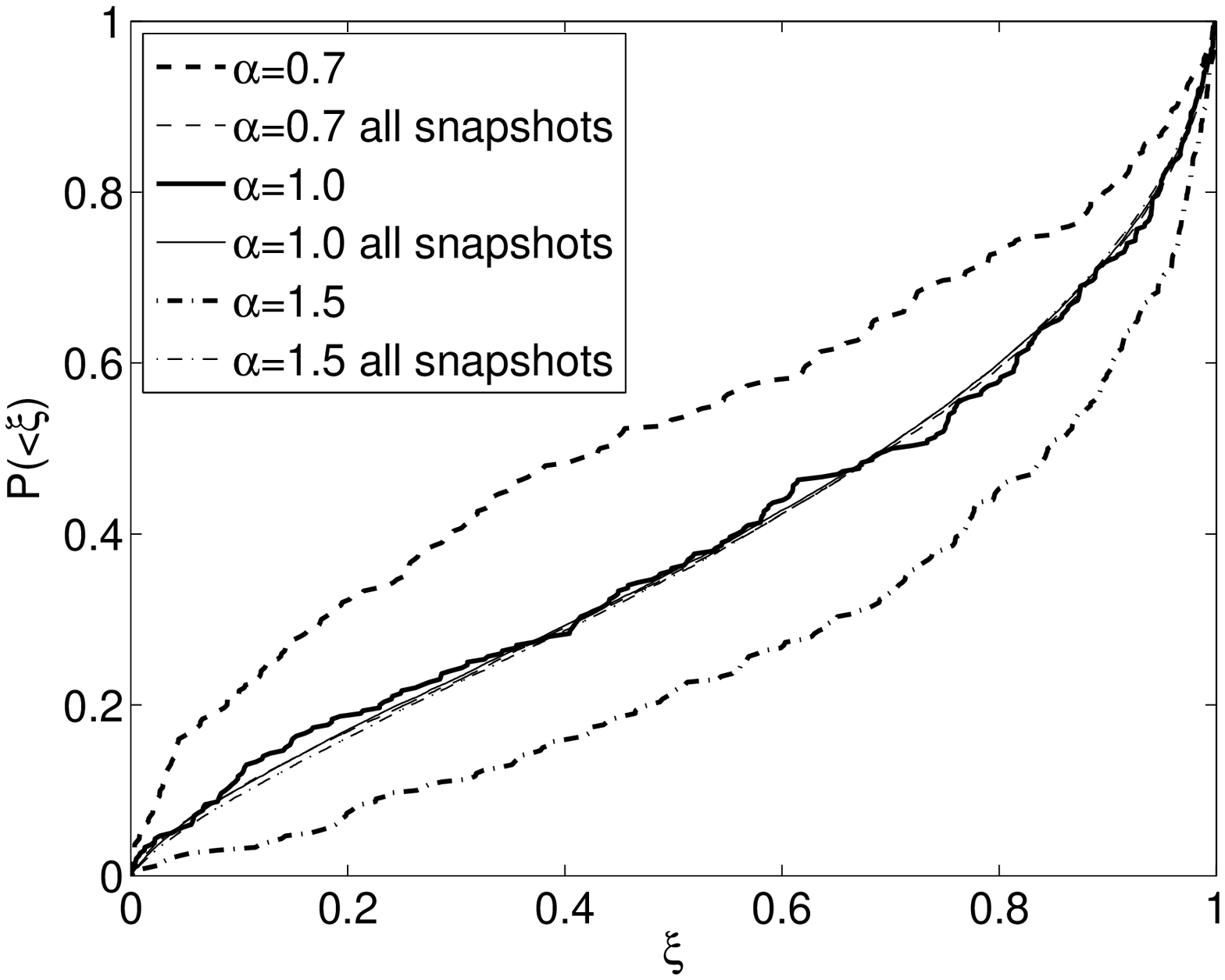}
\includegraphics[   scale=0.4 ,angle=00]{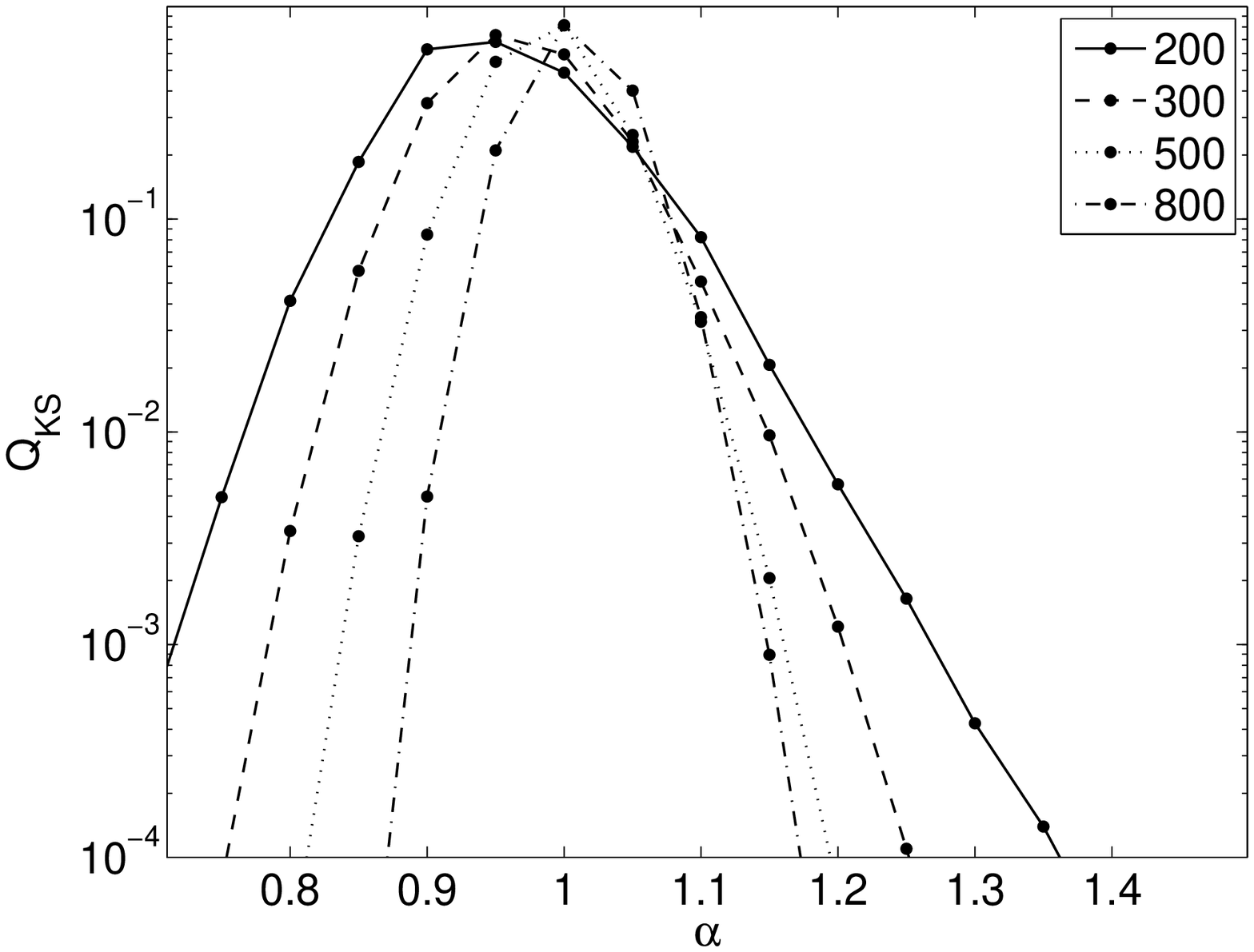}
\vspace{0pt}
\caption{{\it Left:}  CPDFs of $\xi$ for 300 tracers. 
The three thick lines represent ``observed" $P(<\xi)$  computed with 
 $\rmin$ and $\rmax$  corresponding to $\alpha=0.7$, 1 and $1.5$, 
 as indicated in the figure.  The nearly overlapping three  smooth thin lines 
 represent the model  $\Pm(<\xi)$ corresponding to the same choice of  $\alpha$ values. {\it Right:} The confidence level $\Qks$ 
 as a function of $\alpha$. }
\label{fig:pxi}
\end{figure*}

 We now explore  the CPDFs of $\xi$.  The CPDF  $P(<\xi)$
 computed from the   ``observed" particle positions in the mock 
catalog depend on the assumed $\alpha$ through the pericenters, $\rmin$, and apocenters, 
$\rmax$, of the orbits. This is in contrast to the observed $P(<r)$ which is completely independent of 
$\alpha$. 
The three thick lines in  the left panel of Fig.~\ref{fig:pxi} are 
the CPDFs $P(<\xi)$, computed from the ``observed"  positions of 300 particles, for 
$\alpha=0.7$, 1 and 1.5, as indicated in the figure. The three nearly overlapping thin dashed, thin solid and 
thin dot-dashed lines show $\Pm(<\xi)  $, respectively,  for  these three $\alpha$ values. 
The dependence of $\Pm(<\xi)$ on $\alpha$ is very weak, in contrast to $\Pm(<r)$ which is very 
sensitive to
the assumed $\alpha$. This weak dependence is    not entirely  
unexpected  since $\xi$ eliminate the overall length scale of the problem.
 For $\alpha=0.7$, the  difference between $\Pm(<\xi)$ (thin dashed line) and 
 $P(<\xi)$ (thick dashed) implies 
that most of the ``observed" positions are closer to the pericenters than the 
 positions in 
the  snapshots.
For $\alpha=1.5$  (thin and thick dot-dashed lines), the ``observed" positions 
are closer to the apocenters 
than the positions 
in the snapshots.
For the correct value $\alpha =1$, 
the distributions $\Pm(<\xi)$ (thin solid) and $P(<\xi)$ (thick solid) 
appear to be consistent.
To test whether the difference between $\Pm(<\xi)$ and $P(<\xi)$ is large enough to rule an 
assumed $\alpha$  we show in
Fig.~\ref{fig:pxi}  the corresponding  quantity  $Q_{_{\rm KS}}$ as a function of $\alpha$ for 
$n=200$, 300, 500 and 800,  as indicated in the figure.  
 As is the case for the CPDFs of $r$,  the  value $\alpha=1$ 
is never ``rejected" 
at more than the $1\sigma $ level. 
All curves are narrower than the corresponding curves  in the right panel of 
Fig.~\ref{fig:pr}, implying  a better ability to constrain $\alpha$ with the distribution of
 $\xi$ than of $r$.
For $n=200$, 
 deviations of  more than  $20\%$ from $\alpha=1$ 
 are rejected at more than the  $2\sigma$ ($\Qks<0.045$) 
 level.
 For a given confidence level, the range of $\alpha$ (i.e. confidence interval) 
 constrained  by the distribution of $\xi$ is narrower by a factor of 2 than the range 
 constrained by distribution of $r$.

\begin{figure*} 
\centering
\includegraphics[   scale=0.4 ,angle=00]{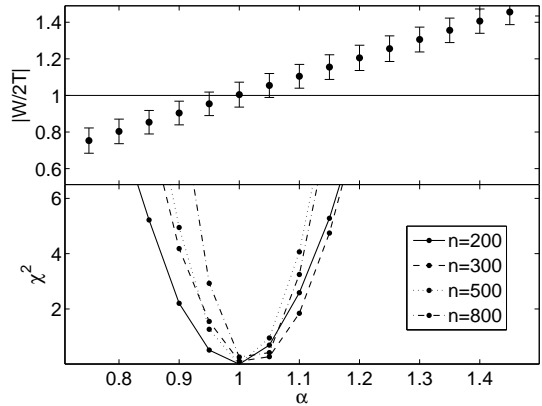}
\vspace{0pt}
\caption{{\it Left:} The filled circles show the ratio, $|W/2T|$, versus $\alpha$ from the observed 
distances and velocities of 200 mock tracers at $\ti$. The error-bars show the r.m.s. scatter, $\sigma$,
computed from  snapshots at $t>\ti$. The horizontal line indicates  the virial theorem value for 
$\overline W/2\overline T$.    {\it Right:}  The quantity $\chi^2= [|W/2T|-1]^2/\sigma^2$ versus $\alpha$.}
\label{fig:vir}
\end{figure*}

The statistic $|W/2T|$ computed form the mock observations with $n=200$
is presented as a function of $\alpha$  by the filled circles in the left panel of
 Fig.~\ref{fig:vir}. 
Here $T$ is computed with the 
``observed" velocities and 
the dependence on $\alpha$ is entirely due to the $W$ term (see Eq.~\ref{eq:vir}).
For each $\alpha$, the error-bars show the r.m.s scatter, $\tilde \sigma$,  of $|W/2T|$ estimated from snapshots
generated with that $\alpha$.
As a consistency check we confirm that the  departure 
 from unity of the ratio $\overline W/2\overline T$, obtained
 by averaging $W$ and $T$ over the snapshots, is completely insignificant compared to 
 the $1\sigma $ scatter.
 The deviation from unity of  $|W/2T|$ computed from the mock observations 
 is shown in the right panel of the Fig.~\ref{fig:vir} as $\chi^2=[|W/2T|-1]^2/{\tilde \sigma}^2$. 
For $n=200$, this statistic constrains $\alpha$ to better than $15\%$ at the $2\sigma$ 
confidence level. This is better than 
the $N$ statistic which constrains $\alpha$ to better than $25\%$ at the same confidence level. 
It also fares slightly better than  the  $\xi$ distribution.

\subsection{Effect of measurement  errors}
So far we have assumed zero errors in the phase space coordinates. We  
present here only  
partial tests of  the method when applied 
on noisy data.  The amplitude of the  errors depends on the distance of tracers from 
an observer at the solar position rather than the galactocentric distances. We  
assign a 15\%  
error in the distance of a  tracer from  an observer present in the Galactic  disk  at
 8$\rm kpc$ from the 
Galactic center. For GAIA  a 10\% parallax distance error corresponds to objects of 
15 mag at 
$10\rm kpc$ from the 
observer. Errors in parallax distances  scale quadratically with true distance and 
become very large at 
distances of tens of kpcs even for GAIA and SIM. 
Here we assume that distances of far away tracers 
are determined by other means so that a linear scaling of the errors is maintained. 
We also perturb the radial velocities and  proper motions with errors that scale linearly 
with distance.
The amplitude velocity errors is assumed equal in all three directions and 
is normalized to a r.m.s value of $10\kms$ at a distance of $20 \kpc$. For comparison, at 15 mag, GAIA 
after 5 years of 
operation  will provide radial motions within 
an accuracy of $10-15\kms$  and proper motions within an accuracy of $1\kms $ at a $20 \kpc$.  

\begin{figure*} 
\centering
\includegraphics[   scale=0.4 ,angle=00]{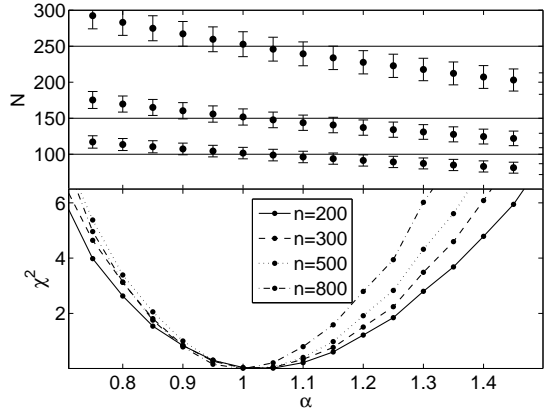}
\vspace{0pt}
\caption{The same as Fig.~\ref{fig:N}, but with degraded mock observations by 
random errors as described in the text.}
\label{fig:Nf}
\end{figure*}

\begin{figure*} 
\centering
\includegraphics[   scale=0.4 ,angle=00]{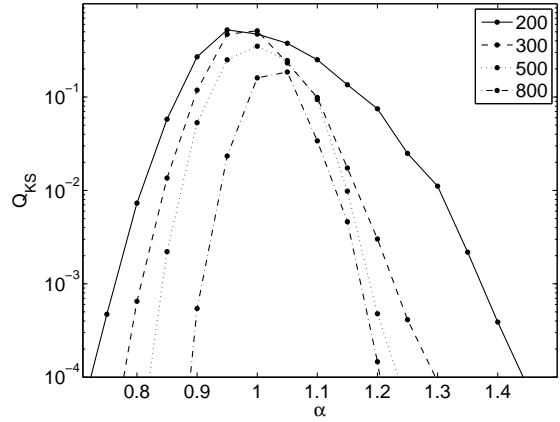}
\vspace{0pt}
\caption{The same as the right panel of Fig.~\ref{fig:pxi}, but with degraded mock observations.}
\label{fig:Qksf}
\end{figure*}

\begin{figure*} 
\centering
\includegraphics[   scale=0.4 ,angle=00]{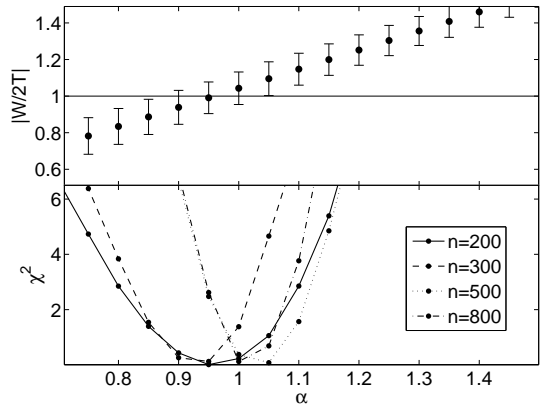}
\vspace{0pt}
\caption{The same as Fig.~\ref{fig:vir}, but with degraded mock observations. }
\label{fig:virf}
\end{figure*}

We do not show results for $P(<r)$ computed with errors included as it  yields less significant 
constraints than 
the remaining statistics.
The results for $N$ and $\xi$ are shown  in Fig.~\ref{fig:Nf} and Fig.~\ref{fig:Qksf}, 
respectively.
For $n=200$ and 300, there is little difference between the curves of $\chi^2$ and $\Qks$ in 
these figures and 
 in Figs.~\ref{fig:N}  \& \ref{fig:pxi} corresponding to zero errors. For larger $n$ the
  effect of the errors
 is more pronounced, especially in $\Qks$. 
 The reason for this behavior is that sampling errors resulting from the finite number of 
 tracers are dominant over measurement errors for the smaller values of $n$. 
 As $n$  increases, measurement errors become more pronounced resulting in the 
 significant reduction of the value of $\Qks$ at $\alpha=1$ for $n=800$. The $N$ statistic
 seems to be more resilient to measurement errors than  the $\xi$ distribution. 
The  virial ratio $|W/2T|$ computed with noisy mock data is shown in Fig.~\ref{fig:virf} with
 the same notation as 
Fig.~\ref{fig:vir} corresponding to results with zero measurement errors.  In computing $T$, the mean of 
non-vanishing   quadratic terms due to  velocity errors have been removed. 
Measurement errors seem to affect the ratio more than  the other statistics. Still, this ratio 
constraints $\alpha$ slightly better than the other statistics.

Overall, measurement  errors of the amplitude we consider here have not degraded the ability 
of the method at constraining $\alpha$. 
Therefore, the method does not require unrealistically accurate data. 

\subsection{Effect of gradually  growing disk}
\begin{figure} 
\centering
\includegraphics[   scale=0.4 ,angle=00]{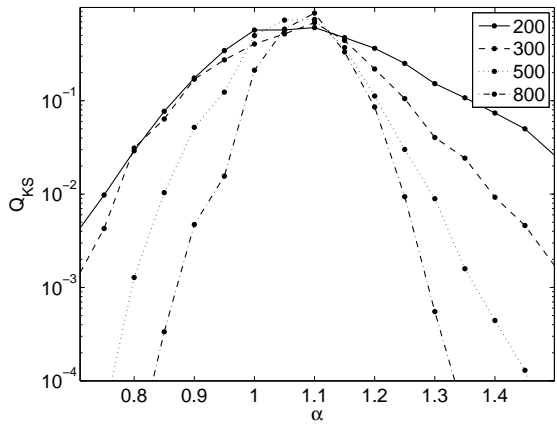}
\vspace{0pt}
\caption{The  confidence level $\Qks$ as a function of $\alpha$, assuming  a gradually growing disk (see text). }
\label{fig:adia}
\end{figure}

The method assumes a constant gravitational potential.  However, the 
Galactic disk may have grown substantially in the last 8Gyr or so. 
Here we  check whether this gradual growth seriously hampers the application of the method.  
We solve for the orbits of tracers assuming that 
the mass distribution of the disk grows like $t^{0.3}$.
The method is then applied to the resultant distribution of tracers in phase space assuming 
a constant disk.  In Fig.~\ref{fig:adia} we show the confidence level $\Qks$ as a function of $\alpha$
 where in this 
case the mass in the disk is assumed to be known and $\alpha$ describes the 
ratio of the assumed $\vh$ to the actual value. 
\section{Concluding Remarks}
\label{sec:Conclude}

The ergodic method presented here  requires a parametric functional form for the Gravitational potential,
 but does not impose  any special symmetry 
on the mass distribution. 
The method assumes that the observations of a class of tracers are spatially complete. 
Tracers with observed distances smaller than  $d_0$, could be present 
beyond $d_0$ in snapshots at later times. Therefore, 
Observational selection against tracers at distances $>d_0$ will make the statistical comparison between
observations and snapshots at later times extremely difficult.
The completeness is 
not too demanding a condition for tracers like globular clusters and Galactic satellites 
for which future observations
should be  
accurate enough for quite  large Galactic  distances.  


We have presented only partial testing of the method, with only a one parameter family for
the form of the Galactic potential. Method is able to constrain the parameter to a good accuracy 
 with measurements errors that are even   larger than those expected to be  achieved by future data. 
The tests show that the method could provide unbiased constraints on the Gravitational potential, 
but  a more elaborate testing which includes 
a more realistic treatment of the errors should be done. Observations will  likely assign distance 
and velocity 
 measurement errors to tracers  on an individual basis. 
 Therefore,  random errors and systematic biases tailored to the specific sample of tracers used by
  the method 
 could be determined robustly.  
 
  The accuracy of the method is mainly limited by the number of tracers. 
  The most obvious tracers  are globular clusters and Galactic satellites. Our
   Galaxy includes 
  158 known globular clusters, and
 23 known satellites \cite[e.g.][]{SG07}.
   Distance measurements of RR Lyrae stars from  
their period-luminosity relation will be greatly improved by  GAIA and SIM   calibration of the 
zero point using a nearby 
sample of these stars.
Therefore, luminous halo RR Lyrae stars could significantly enlarge the sample of tracers. 

The method requires a   system of tracers in dynamical equilibrium in the 
current Galactic potential. A pre-requist for dynamical equilibrium is that any recent changes in the 
Galactic potential  must have occurred on a time scale longer that the dynamical 
time of the system of tracers. 
Spectroscopic studies stars in the Galaxy do not present evidence for 
  substantial mergers in the last 8Gyr \cite[]{GWN,Helmi}, implying a  nearly static 
  gravitational potential.   We have demonstrated that  adiabatic growth of the Galactic disk is not expected to pose a problem for the implementation of the method.
 The effect of any deviation from that should be modeled. 

 One issue which is exclusive to using disk stars as tracers is  the effect of transient perturbations
 on the dynamics of those stars \cite[e.g.][]{FAM05,QUIL}. 
  However,  the transient effects
cause velocity perturbations at the level of ~10 km/s which is much smaller that 
the total velocities of tracers.  Observational uncertainties are larger than that and do not
seem to cause significant biases in the results as indicated by the tests described above.
  Another issue to be considered is halo substructure  which, in principle, could 
  act as a stochastic component in the gravitational potential. Such a component is 
  extremely difficult to model in the method proposed here. We offer the following argument demonstrating 
  that substructure should not have an 
  important effect on the long term dynamics of tracers. 
  In the limit of fast encounters, a tracer passing a substructure at distance $b$ will 
  change its velocity by $V_1\approx g(b)b/V$ where $V$ is the relative velocity 
  and $g\approx G M/b^2$ is the gravitational force field of the substructure assuming that $b$ is larger than its 
  tidal radius. Performing the usual summing in quadratures over encounters occurring in one orbital time
  we get that the  r.m.s change
  $<V_1^2>^{1/2}\approx M\sqrt{\bar n R}/V$ where $\bar n$ is  the number density of 
  substructures and $R$ is the distance travelled by the tracer in one orbital time. 
  Taking $V^2=G M_{_{\rm MW}}(R)/R$ with $M_{_{\rm MW}}(R)$  the mass of the Galactic halo within
  radius $R$, we get the condition $\sim M_{_{\rm MW}} \sqrt{\bar n R^3}>M_{_{\rm MW}}$ for having $<V_1^2>^{1/2}\sim V$.
 This means that neither single encounters nor collective  stochastic effects  can dominate the  long term evolution of tracers even if the fraction of mass in substructures is large which is 
 contrary to  recent simulations \cite[e.g.][]{Colombi} which show that the smooth component 
 greatly dominates the mass of Galactic size halos.

 When details of future data from SIM and GAIA become available, all robust information
 about the distribution of baryons in the Galaxy should be used \cite[e.g.][]{ROBIN03}
 in order to place tight constraints on the Galactic dark matter.
 The validity of the method should be tested with mock data that match the observations 
 as much as possible and with the best possible available Galactic models.

\section{Acknowledgments}
The author wishes to thank an anonymous referee for comments which helped 
improve the paper. 
This work is supported by the German-Israeli Foundation for 
Research and Development and by the Asher Space Research
Institute.  This research was supported by S. Langberg Research Fund.
\bibliography{refs}

\end{document}